\documentclass[conference]{IEEEtran}
\IEEEoverridecommandlockouts
\usepackage{cite}
\usepackage{amsmath,amssymb,amsfonts}
\usepackage{algorithmic}
\usepackage[pdftex]{graphicx}
\usepackage{textcomp}
\usepackage{xcolor}
\usepackage{comment}
\def\BibTeX{{\rm B\kern-.05em{\sc i\kern-.025em b}\kern-.08em
    T\kern-.1667em\lower.7ex\hbox{E}\kern-.125emX}}
\begin{document}

\title{Effective Capacity Analysis of Joint Near and Far-Field Communication in 6G URLLC Networks\\

\author{
\IEEEauthorblockN{
Humera Hameed\IEEEauthorrefmark{1}, Waqas Aman\IEEEauthorrefmark{2}, Muhammad Mahboob Ur Rahman\IEEEauthorrefmark{1}, Ali Arshad Nasir \IEEEauthorrefmark{3} }

\IEEEauthorblockA{\IEEEauthorrefmark{1}Electrical engineering department, Information Technology University (ITU), Lahore 54000, Pakistan\\ \IEEEauthorrefmark{2}College of Science and Engineering, Hamad Bin Khalifa University (HBKU), Doha, Qatar \\
\IEEEauthorrefmark{3}Department of Electrical engineering, King Fahd University of Petroleum and Minerals (KFUPM), Dhahran, Saudi Arabia\\
\IEEEauthorrefmark{1}\{phdee17001, mahboob.rahman\}@itu.edu.pk, \IEEEauthorrefmark{2}waqasaman87@gmail.com, \IEEEauthorrefmark{3}anasir@kfupm.edu.sa
} 
}
% \author{\IEEEauthorblockN{1\textsuperscript{st} Humera Hameed}
% \IEEEauthorblockA{\textit{Department of Electrical Engineering} \\
% \textit{Information Technology University}\\
% Lahore, Pakistan \\
% \and
% \IEEEauthorblockN{2\textsuperscript{nd} Waqas Aman}
% \IEEEauthorblockA{\textit{College of Science and Engineering} \\
% \textit{ Hamad Bin Khalifa University}\\
% Doha, Qatar \\
% waman@hbku.edu.qa}
% \and
% \IEEEauthorblockN{3\textsuperscript{rd} Muhammad Mahboob Ur Rahman}
% \IEEEauthorblockA{\textit{Department of Electrical Engineering} \\
% \textit{Information Technology University}\\
% Lahore, Pakistan \\
% \and
% \IEEEauthorblockN{4\textsuperscript{th} Ali Arshad Nasir}
}
\maketitle

\begin{abstract}
The emergence of 6G networks enables simultaneous near-field and far-field communications through extremely large antenna arrays and high carrier frequencies. While these regimes enhance spatial multiplexing and link capacity, their coexistence poses new challenges in ensuring quality-of-service (QoS) guarantees for delay-sensitive applications. This paper presents an effective capacity (EC) analysis framework that jointly models near- and far-field communication regimes under distance estimation uncertainty. The user location is modeled as a random variable spanning both propagation regions, and tractable closed-form expression for the EC is derived to quantify delay performance. Numerical results illustrate the impact of estimation variance, QoS exponent, far-field boundary and near-field boundary (Fraunhofer distance) on EC performance. 
\end{abstract}


\section{Introduction}

%The global rollout of fifth-generation (5G) networks has catalyzed intensive efforts toward the development of sixth-generation (6G) wireless technologies. 
The 6G vision aims to support emerging paradigms such as integrated sensing and communications (ISAC), extended reality (XR), holographic telepresence, and intelligent human–machine interaction, all of which demand extreme performance in latency, reliability, and data rate. Achieving these capabilities requires the exploitation of higher frequency bands—spanning the centimeter-wave, millimeter-wave (mmWave), and terahertz (THz) spectra—where much broader bandwidths are available. However, signals in these bands experience severe propagation loss and weaker diffraction compared to sub-6 GHz systems, making large-scale antenna arrays essential for compensating path loss and enabling fine spatial resolution. This evolution has led to the concept of extremely large aperture arrays (ELAAs), comprising hundreds or thousands of antennas that deliver exceptional beamforming gain, enhanced spatial selectivity, and improved spectral efficiency \cite{carvalho2020non}.

The adoption of ELAAs fundamentally shifts the propagation regime from the far field (FF) to the near field (NF), where classical modeling assumptions no longer hold. The electromagnetic (EM) field radiated by an antenna array can be divided into the reactive NF, radiative NF, and far field regions. Since the reactive NF is confined to non-propagating fields near the antenna, the radiative NF is most relevant to communication and sensing. In this region, wavefronts are spherical, and the path-length variations across the array aperture become significant. Thus, the spherical wavefront model (SWM) accurately describes NF propagation, albeit at increased computational cost. Conversely, the planar wavefront model (PWM), widely used in 5G, assumes planar waves and ignores higher-order phase terms, simplifying signal processing. As array apertures grow in ELAAs, the Rayleigh distance—the boundary between NF and FF—can extend from meters to hundreds of meters, causing many users to operate within the NF region.

The NF regime introduces distinct physical and modeling characteristics with major design implications. NF channels exhibit spatial non-stationarity, where different subarrays may observe different scatterers, leading to diversity in angle and delay profiles. The channel rank and capacity may increase due to the additional spatial degrees of freedom inherent in the SWM \cite{jiang2005spherical}. Moreover, path-loss scaling differs from that of FF channels, necessitating revised models for beam management, power control, and channel estimation \cite{carvalho2020non}, \cite{liu2023near,chen20246g,elzanaty2024near}. Hence, accurately defining and characterizing the NF–FF boundary is vital for realistic channel modeling, for performance optimization in 6G systems, and for the design of optimal beamforming.
That is, the traditional FF beamforming depends only on the angle of arrival (AoA) and often uses discrete Fourier transform-based codebooks. In contrast, NF beamforming jointly depends on distance and angle, leveraging the spherical curvature of wavefronts. This joint control not only improves communication performance but also enables precise localization and environmental sensing, key enablers of ISAC systems \cite{chen20246g}, \cite{elzanaty2024near}.

Simultaneously, the rise of ultra-reliable low-latency communication (URLLC) services in 6G introduces stringent delay and reliability constraints that extend beyond physical-layer considerations. Emerging use cases such as industrial automation, teleportation, and remote surgery demand predictable and bounded end-to-end latency, including that of wireless backhaul and front-haul segments. In this regard, effective capacity (EC) theory provides a mathematically rigorous framework to characterize the maximum constant data arrival rate that a communication system can sustain under statistical quality-of-service (QoS) constraints \cite{Wu:TWC:2003}. By incorporating channel fading statistics and delay violation probabilities into a unified model, EC captures the interplay between reliability, latency, and throughput, offering a more holistic view of system performance \cite{Wang:ICST:2023}. Recent studies have extended the EC framework to a variety of wireless technologies, including cognitive radio \cite{Anwar:TVT:2016}, mmWave and THz communication \cite{Poor:ITC:2018, Han:TWC:2024}, reconfigurable intelligent surfaces (RIS) \cite{aman2021effective, shah2023effective}, and hybrid THz/FSO systems \cite{Aman:WCL:2025}. Additionally, the impact of system-level mechanisms—such as mode selection \cite{WShah:WCL:2019}, physical-layer authentication \cite{waqas:ICC:2020}, and channel state information (CSI) imperfections—on EC has been systematically investigated. {Note that all the notable existing studies on EC analysis have focused solely on a single propagation regime, typically the FF region. To the best of our knowledge, no prior work has jointly modeled both NF and FF regimes within a unified framework to investigate the impact of regime transitions on QoS-constrained performance.}

In light of the above, extending EC analysis to NF and FF communication regimes constitutes an open and timely research direction. The presence of spherical wave-fronts, spatial non-stationarity, and position-dependent channel correlations in the NF profoundly alters the statistical structure of instantaneous rates, rendering conventional FF-based EC formulations inadequate. A rigorous EC analysis for NF–FF systems can provide new insights into how physical-layer characteristics—such as array geometry, user distance, and wavefront curvature—affect higher-layer QoS metrics. Such cross-layer understanding is essential for designing next-generation ISAC and URLLC systems that are not only spectrally and spatially efficient but also delay-aware and reliability-guaranteed under realistic NF propagation conditions.

\textbf{Contributions.}  
The main contributions of this paper are summarized as follows:
\begin{itemize}
    \item A simple distance-based hypothesis test is proposed to distinguish between NF and FF propagation regimes.
    \item A tractable closed-form expression of the  EC is derived for joint NF and FF regimes communication, accounting for regime misclassification and rate outage events.
    \item Extensive numerical simulations are provided to study the impact of ranging uncertainty, FF boundary, Fraunhofer distance and QoS constraints.
\end{itemize}

\textbf{Outline.}  
The remainder of this paper is organized as follows. Section~II introduces the system model and channel formulations for NF and FF regions. Section~III presents the distance-based regime classification method and its error analysis. Section~IV develops the effective capacity framework using a Markov-state model. Simulation results are given in Section~V, followed by concluding remarks in Section~VI.

\section{System Model}
We consider a MIMO communication system employing continuous-aperture uniform linear arrays (ULAs) at both the base station (BS) and the user terminal (UT) (see Fig. 1). The physical lengths of the transmit and receive apertures are denoted by $L_t$ and $L_r$, respectively, with $L_t \ge L_r$ assumed without loss of generality. The carrier wavelength is $\lambda$, the transmit power is $P$, and the receiver noise power spectral density is $N_0$, yielding a signal-to-noise ratio (SNR) of $\rho = \tfrac{P}{N_0}$. The UT is assumed to be uniformly distributed over a circular cell area, representing a two-dimensional spatially uniform user distribution. A downlink transmission is considered for EC at BS, where the communication is subject to QoS constraints. The BS estimates the UT distance $\hat{d}$ and classifies the propagation regime—NF or FF—accordingly. Based on this classification, the BS schedules the transmission rate from its data queue using the appropriate Shannon capacity expression defined later.

\begin{figure}[htb!]
    \centering
    \includegraphics[width=\linewidth]{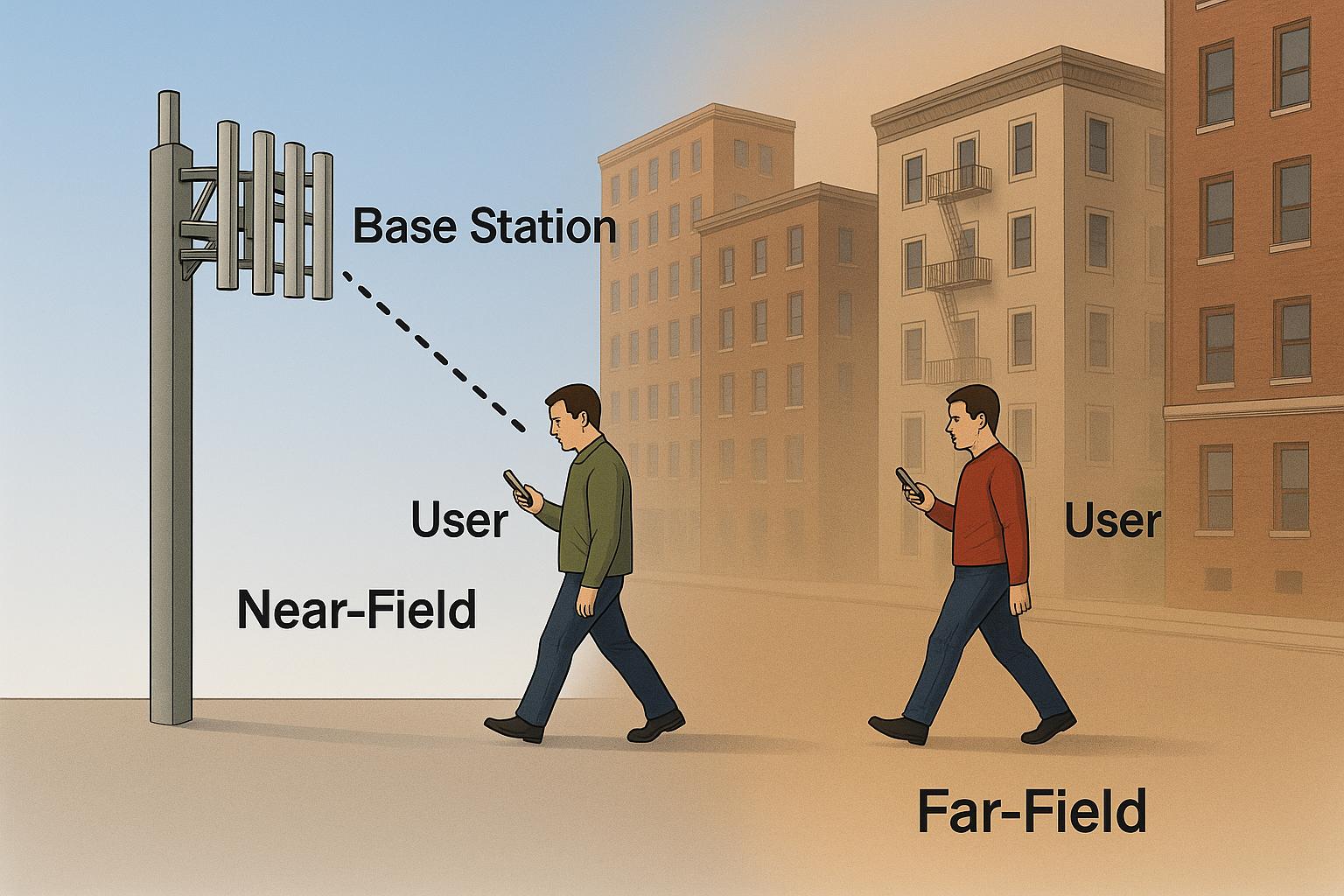}
    \caption{Illustration of near-field and far-field propagation regimes in a  continuous-aperture MIMO system.}
    \label{fig:placeholder}
\end{figure}

The transition between the NF and FF propagation regimes is given by the Rayleigh (or Fraunhofer) distance \cite{Sun2025Nearfield}
\begin{equation}
d_F = \frac{2 L_t L_r}{\lambda},
\end{equation}
which depends on the aperture lengths and wavelength.
\subsubsection*{Near-Field (NF) Channel}
In the NF region ($d < d_F$), the electromagnetic wavefront is
spherical, leading to spatial non-stationarity across the aperture.
Following the continuous-aperture MIMO formulation, the capacity conditioned on
distance $d$ is expressed as \cite{Xie2023NearField}
\begin{equation}
\label{eq:near_capacity}
    C_{\mathrm{n}}(d)
    = u(d)
      \log_2\!\left(
      1 +
      \frac{c_0
      \!\left(
      \tfrac{L_t L_r \lambda}{d^{3}}
      - \tfrac{\lambda^2}{4 d^{2}}
      \right)^{\!2}}
      {\!\left(\tfrac{2 L_t L_r}{d^{2}} - v(d)\right)^{\!3}}
      \rho
      \right),
\end{equation}
where $c_0$ is a constant that depends on the aperture normalization and antenna
radiation pattern. The auxiliary functions $u(d)$ and $v(d)$ represent array
coupling and phase compensation effects, respectively, and are defined as
\begin{align}
    u(d)
    &= \frac{\big(2 L_t L_r - d^{2} v(d)\big)^{2}}
            {L_t L_r \lambda d - \lambda^{2} d^{2}/4}, \\[3pt]
    v(d)
    &= \ln\!\left(
       \frac{(L_t + L_r)^2 + 4 d^2}
            {(L_t - L_r)^2 + 4 d^2}
       \right).
\end{align}
This formulation captures the rapid spatial variation of the channel gain and
phase coupling as the UT moves within the Fresnel region.

\subsubsection*{Far-Field (FF) Channel}
For $d \ge d_F$, the electromagnetic wavefront becomes approximately planar,
and the array response reduces to a directional gain. The capacity thus
simplifies to
\begin{equation}
\label{eq:far_capacity}
    C_{\mathrm{f}}(d)
    = \log_2\!\left(
      1 +
      \frac{c_0 L_t L_r}{d^{2}} \rho
      \right).
\end{equation}
This expression follows the conventional Friis-based MIMO capacity, where
path-loss scales inversely with the square of distance. The NF
expression in \eqref{eq:near_capacity} generalizes this by incorporating
spherical wavefront and spatial-focusing effects, which become dominant in
large-aperture transceivers.

\section{Distance-Based Regime Classification}
The BS aims to determine whether a UT lies in the NF ($d < d_F$) or
FF ($d \ge d_F$) region based on the estimated distance $\hat d$,
obtained from time-of-arrival (ToA) measurements (the details of the ToA estimation and corresponding estimation uncertainty is provided in Appendix):
\begin{equation}
    \hat d = d + \varepsilon,
    \qquad
    \varepsilon \sim \mathcal{N}(0,\sigma_d^2),
\end{equation}
where $\varepsilon$ denotes the zero-mean Gaussian ranging noise with variance
$\sigma_d^2$.
This problem of classification can be formulated as a binary hypothesis test:
\begin{equation}
    \begin{cases}
        \mathcal{H}_0: \hat{d} < d_F, & \text{(NF)}\\[3pt]
        \mathcal{H}_1: \hat{d} \ge d_F, & \text{(FF)}
    \end{cases}
\end{equation}
where  $\mathcal{H}_0$, and $\mathcal{H}_1$ indicate UT is in NF and FF, respectively. 
Then, the decision rule is
\begin{equation}
    \hat d
    \mathop{\gtrless}_{\mathcal{H}_1}^{\mathcal{H}_0}
    d_F.
\end{equation}
\subsubsection*{Classification errors with physical truncation (\(\hat d\ge 0\))}
Under truncation the conditional PDF of \(\hat d\) (given the true distance \(d\))
is the Gaussian density truncated below at 0, and the renormalizing factor is
\[
\Pr(\hat d \ge 0 \mid d)=1-\Phi\!\Big(-\frac{d}{\sigma_d}\Big)=\Phi\!\Big(\frac{d}{\sigma_d}\Big),
\]
where $\Phi(\cdot)$ denotes the standard normal CDF.
The conditional False–Far (Type I) error under truncation is
For \(d<d_F\),
\[
\begin{aligned}
P_{\mathrm{FF}}(d) 
&= \Pr(\hat d \ge d_F \mid d,\,\hat d\ge0) \\
&= \frac{\Pr(\hat d \ge d_F \mid d)}{\Pr(\hat d \ge 0 \mid d)}
= \frac{1-\Phi\!\big(\tfrac{d_F-d}{\sigma_d}\big)}{\Phi\!\big(\tfrac{d}{\sigma_d}\big)} 
= \frac{Q\!\big(\tfrac{d_F-d}{\sigma_d}\big)}{\Phi\!\big(\tfrac{d}{\sigma_d}\big)},
\end{aligned}
\]
where $Q(.)$ denotes standard Q-function.  
Similarly, the conditional False–Near (Type II) error under truncation is
For \(d\ge d_F\),
\[
\begin{aligned}
P_{\mathrm{FN}}(d)
&= \Pr(\hat d < d_F \mid d,\,\hat d\ge0) \\
&= \frac{\Pr(0 \le \hat d < d_F \mid d)}{\Pr(\hat d \ge 0 \mid d)} 
= \frac{\Phi\!\big(\tfrac{d_F-d}{\sigma_d}\big)-\Phi\!\big(-\tfrac{d}{\sigma_d}\big)}
        {\Phi\!\big(\tfrac{d}{\sigma_d}\big)}.
\end{aligned}
\]
\subsubsection*{Unconditional misclassification probabilities}
Assuming UT is uniformly distributed in area over the annulus \(d\in[d_{\min},d_{\max}]\), so the marginal pdf of the true distance is
\[
f_D(d)=\frac{2d}{d_{\max}^2-d_{\min}^2},\qquad d_{\min}\le d\le d_{\max}.
\]
\paragraph{False–Far (Type I):}
For \(d\in[d_{\min},d_F)\), the unconditional false–far probability (averaged over NF distances) is
\begin{equation}
P_{\mathrm{FF}}
= \int_{d_{\min}}^{d_F}
\frac{Q\!\big(\tfrac{d_F-d}{\sigma_d}\big)}
     {\Phi\!\big(\tfrac{d}{\sigma_d}\big)}
\;\frac{2d}{d_{\max}^2-d_{\min}^2}\, \mathrm{d}d.
\label{eq:PFF_trunc_uncond}
\end{equation}

\paragraph{False–Near (Type II):}
The unconditional false–near probability (averaged over FF distances) is
\begin{equation}
P_{\mathrm{FN}}
= \int_{d_F}^{d_{\max}}
\frac{\Phi\!\big(\tfrac{d_F-d}{\sigma_d}\big)-\Phi\!\big(-\tfrac{d}{\sigma_d}\big)}
     {\Phi\!\big(\tfrac{d}{\sigma_d}\big)}
\;\frac{2d}{d_{\max}^2-d_{\min}^2}\, \mathrm{d}d.
\label{eq:PFN_trunc_uncond}
\end{equation}

\section{Effective Capacity (EC) Analysis}
%\begin{definition}
EC is the log moment-generating function
(MGF) of the cumulative service process $\sum_{k=1}^t s(k)$ at steady state
\cite{Wu:TWC:2003}:
\begin{equation}
\label{eq:ECStandard}
\text{EC} 
= -\lim_{t \to \infty} \frac{1}{\theta t}
\ln \Big\{ \mathbf{E} \big[ e^{-\theta \sum_{k=1}^t s(k)} \big]\Big\},
\end{equation}
where $k$ is the index used for denoting discrete time-slot, $t$ index
 is used for real time, $\mathbf{E}[\cdot]$ denotes the expectation operator, and $\theta$ is the
QoS exponent that captures delay sensitivity: $\theta \to 0$ implies
delay-tolerant communication, whereas $\theta \to \infty$ implies
delay-intolerant communication.
%\end{definition}

\subsection{Markov Chain Representation}
Due to UT position, the distance $d$ varies across time slots, causing the user to transition between NF and FF regimes, and the noisy estimates of the distance make the classification process prone to ranging errors. This can be modeled using a Markov chain with eight possible states\footnote{We omit the time index $k$ for the sake of notational brevity.}:
\begin{itemize}
    \item \textbf{State 1 ($S_1$):} $\mathcal{H}_0|\mathcal{H}_0$ and $R(\hat d) \le C_n(d)$ (correctly classified NF and reliable transmission).
    \item \textbf{State 2 ($S_2$):} $\mathcal{H}_0|\mathcal{H}_0$ and $R(\hat d) > C_n(d)$ (correctly classified NF but transmission outage).
    \item \textbf{State 3 ($S_3$):} $\mathcal{H}_1|\mathcal{H}_0$ and $R(\hat d) \le C_n(d)$ (NF UT misclassified as FF but reliable transmission).
    \item \textbf{State 4 ($S_4$):} $\mathcal{H}_1|\mathcal{H}_0$ and $R(\hat d) > C_n(d)$ (NF UT misclassified as FF and in outage). 
    \item \textbf{State 5 ($S_5$):} $\mathcal{H}_0|\mathcal{H}_1$ and $R(\hat d) \le C_f(d)$ (FF UT misclassified as NF but reliable transmission). \color{black}
    \item \textbf{State 6 ($S_6$):} $\mathcal{H}_0|\mathcal{H}_1$ and $R(\hat d) > C_f(d)$ (FF UT misclassified as NF and in outage).
    \item \textbf{State 7 ($S_7$):} $\mathcal{H}_1|\mathcal{H}_1$ and $R(\hat d) \le C_f(d)$ (correctly classified FF and reliable transmission).
    \item \textbf{State 8 ($S_8$):} $\mathcal{H}_1|\mathcal{H}_1$ and $R(\hat d) > C_f(d)$ (correctly classified FF but transmission outage).
\end{itemize}
 Since the ranging errors are assumed
independent across time slots, the chain is memoryless, and each state occurs with a fixed steady-state probability. The transition probabilities between states depend on the classification accuracy and rate reliability, as detailed in the subsequent derivations. Please note that the chosen transmission rate $R(\hat{d})$ on estimated distance $\hat{d}$ is according to the Shannon capacity formulas for NF and FF as given in Section II.

\textbf{State $S_1$} corresponds to a NF UT correctly classified as NF and experiencing reliable transmission:
\begin{equation}
    S_1 = \{\, H_0 \mid H_0,~ R(\hat{d}) \le C_n(d) \,\}
    = \{\, d < d_F,~ \hat{d} < d_F,~ \hat{d} \ge d \,\}.
\end{equation}
Since $R(\hat{d})$ decreases monotonically with $\hat{d}$, the condition $R(\hat{d}) \le C_n(d)$ implies $\hat{d} \geq d$.  
Given $\hat{d}=d+\varepsilon$ with $\varepsilon\sim\mathcal{N}(0,\sigma_d^2)$.
Under truncation the conditional distribution of \(\hat d\) given the true distance \(d\)
is a Gaussian truncated below at \(0\). For \(d\in[d_{\min},d_F)\) the conditional
probability of State \(S_1=\{d\le\hat d<d_F\}\) becomes
\begin{equation}
\Pr(S_1\mid d)
= \Pr(d \le \hat d < d_F \mid \hat d \ge 0)
= \frac{\Phi\!\big(\tfrac{d_F-d}{\sigma_d}\big)-\Phi(0)}
       {\Pr(\hat d \ge 0 \mid d)}.
\label{eq:S1_cond_trunc}
\end{equation}
Since \(\Pr(\hat d \ge 0\mid d)=1-\Phi\!\big(-\tfrac{d}{\sigma_d}\big)=\Phi\!\big(\tfrac{d}{\sigma_d}\big)\),
we may equivalently write $\Pr(S_1\mid d)
= \frac{\Phi\!\big(\tfrac{d_F-d}{\sigma_d}\big)-\Phi(0)}
       {\Phi\!\big(\tfrac{d}{\sigma_d}\big)}$

Averaging over the NF distance PDF
\(f_{D\mid\mathcal H_0}(d)=\dfrac{2d}{d_F^2-d_{\min}^2}\) yields the unconditional probability
\begin{equation}
P_{S_1}
= \frac{1}{4}
\int_{d_{\min}}^{d_F}
\frac{\Phi\!\big(\tfrac{d_F-d}{\sigma_d}\big)-\Phi(0)}
     {\Phi\!\big(\tfrac{d}{\sigma_d}\big)}
\;\frac{2d}{d_F^2-d_{\min}^2}\, \mathrm{d}d.
\label{eq:S1_uncond_trunc}
\end{equation}

%     \begin{equation}
%     \Pr(S_1\mid d) = \Pr(d \le \hat{d} < d_F)
%     = \Phi\!\left(\frac{d_F-d}{\sigma_d}\right) - \Phi(0),
% \end{equation}

% Since $S_1$ occurs only when $d<d_F$, the NF distance PDF conditioned on $\mathcal{H}_0$ is
% \begin{equation}
% f_{D|\mathcal{H}_0}(d)=\frac{2d}{d_F^2 - d_{\min}^2}, \quad d_{\min}\le d\le d_F.
% \end{equation}
% Thus, the unconditional probability is
% \begin{equation}
% P_{S_1} = \frac{\Pi_n}{4}\frac{2}{d_F^2-d_{\min}^2}
%     \int_{d_{\min}}^{d_F} d\!\left[\Phi\!\left(\frac{d_F-d}{\sigma_d}\right) - \tfrac{1}{2}\right]\!dd,
% \end{equation}
% where $\Pi_n = \frac{d_F^2-d_{\min}^2}{d_{\max}^2-d_{\min}^2}$ is the prior probability of being in near field.  
% After substituting $z = (d_F - d)/\sigma_d$, and using 
% $\int_0^a\Phi(z)\,dz = a\Phi(a) + \phi(a) - \tfrac{1}{\sqrt{2\pi}}$, we get
% \begin{equation}
% P_{S_1}
% =\frac{\Pi_n}{4}\eta_n
% \!\left[
% \gamma_n
% \!\left(
% \Phi(\gamma_n)-\tfrac{1}{2}
% \right)
% +\phi(\gamma_n)
% -\tfrac{1}{\sqrt{2\pi}}
% \right]\!,
% \end{equation}
% where $\eta_n=\frac{2\sigma_d d_F}{d_F^2 - d_{\min}^2}$, $\gamma_n=\frac{d_F - d_{\min}}{\sigma_d}$, $\phi(z)=\tfrac{1}{\sqrt{2\pi}}e^{-z^2/2}$.

% \color{blue}
% \subsubsection*{State \(\mathbf{S_1}\) with physical truncation (\(\hat d\ge 0\))}

\color{black}

\textbf{State $S_2$} corresponds to NF UT correctly classification and experiencing outage:
\[
S_2 = \{\, H_0\mid H_0,~ R(\hat{d}) > C_n(d) \,\}
     = \{\, d<d_F,~ \hat{d}<d_F,~ \hat{d}<d \,\}.
\]
 For \(d\in[d_{\min},d_F)\) the conditional
probability of State \(S_2\) becomes
\begin{align}
\Pr(S_2\mid d)
&= \Pr(0 \le \hat d < d \mid \hat d \ge 0) \\
&= \frac{\Phi\!\big(\tfrac{d}{\sigma_d}\big)-\Phi(0)}{\Phi\!\big(\tfrac{d}{\sigma_d}\big)}, \nonumber
\end{align}
the unconditional probability is
\begin{equation}
P_{S_2}
= \frac{1}{4}
\int_{d_{\min}}^{d_F}
\frac{\Phi\!\big(\tfrac{d}{\sigma_d}\big)-\tfrac{1}{2}}
     {\Phi\!\big(\tfrac{d}{\sigma_d}\big)}
\;\frac{2d}{d_F^2-d_{\min}^2}\, \mathrm{d}d.
\label{eq:S2_uncond_trunc}
\end{equation}

\textbf{State $S_3$} corresponds to NF-UT misclassified as FF while the chosen rate remains reliable:
\[
S_3=\{\,H_1\mid H_0,\ R(\hat d)\le C_n(d)\,\} = \{\,d<d_F,\ \hat d\ge d_F , \hat d\ge d\}.
\]

Conditioned on a fixed \(d\in[d_{\min},d_F)\), and enforcing the physical truncation \(\hat d\ge0\), the conditional probability is
\begin{align}
\Pr(S_3\mid d)
&= \Pr(\hat d \ge d_F \mid d,\,\hat d\ge0) \\
&= \frac{\Pr(\hat d \ge d_F \mid d)}{\Pr(\hat d\ge0\mid d)}
= \frac{Q\!\big(\tfrac{d_F-d}{\sigma_d}\big)}{\Phi\!\big(\tfrac{d}{\sigma_d}\big)}. \nonumber
\label{eq:S3_cond_trunc}
\end{align}

the unconditional probability for $S_3$ is
\begin{equation}
P_{S_3} = \frac{1}{4} \int_{d_{\min}}^{d_F}
\frac{Q\!\big(\tfrac{d_F-d}{\sigma_d}\big)}
     {\Phi\!\big(\tfrac{d}{\sigma_d}\big)}
\;\frac{2d}{d_F^2-d_{\min}^2}\,\mathrm{d}d.
\label{eq:S3_uncond_trunc_total}
\end{equation}

\textbf{State $S_4$} corresponds to NF UT misclassified as FF and scheduled rate exceeds true NF capacity.

By the model assumptions, $H_1\mid H_0$ implies
\[
d<d_F,\qquad \hat d \ge d_F,
\]
while the condition $R(\hat d)>C_n(d)$ (capacities decreasing with distance)
is equivalent to $\hat d<d$. These two requirements are mutually exclusive for
$d<d_F$ (since $\hat d\ge d_F>d$), therefore the event is empty \footnote{Note that one can remove states having zero probability, but we mention them for the sake of complete presentation}:
\[
S_4 = \varnothing,
\qquad
P_{S_4}=0.
\]
\textbf{State $S_5$:} Similar to {State $S_4$}, one can check that the events inside State $S_5$ are  mutually exclusive, therefore, $P_{S_5}=0$.

\textbf{State $S_6$} corresponds to classification of FF UT as NF UT, and scheduled rate at BS exceeds the Shannon limit:
\[
S_6 = \{\, d\ge d_F,~ \hat{d}<d_F,~ R(\hat{d})>C_f(d)\,\}.
\]
% State \(S_6\) is
% \[
% S_6=\{\;d\ge d_F,\ \hat d<d_F,\ C_n(\hat d)>C_f(d)\;\}.
% \]
Under the monotonicity assumption, the outage condition
\(R(\hat d)>C_f(d)\) is automatically satisfied for any \(d\ge d_F\) and
\(\hat d<d_F\). Therefore \(S_6\) reduces to the misclassification event
\(\{d\ge d_F,\ \hat d<d_F\}\).
Let the true-distance PDF for FF conditioned on \(\mathcal H_1\) be
\[
f_{D\mid\mathcal H_1}(d)=\frac{2d}{d_{\max}^2-d_F^2},\qquad d_F\le d\le d_{\max}.
\]

 Under truncated Gaussian error (enforce \(\hat d\ge0\)), the conditional probability becomes
\[
\Pr(S_6\mid d)
= \Pr(0\le \hat d < d_F \mid \hat d\ge 0, d)
= \frac{\Phi\!\big(\tfrac{d_F-d}{\sigma_d}\big)-\Phi\!\big(-\tfrac{d}{\sigma_d}\big)}
       {\Phi\!\big(\tfrac{d}{\sigma_d}\big)}.
\]
Hence the unconditional (overall) probability is
\begin{equation}
P_{S_6}
= \frac{1}{4}\int_{d_F}^{d_{\max}}
\frac{\Phi\!\big(\tfrac{d_F-d}{\sigma_d}\big)-\Phi\!\big(-\tfrac{d}{\sigma_d}\big)}
     {\Phi\!\big(\tfrac{d}{\sigma_d}\big)}\;
\frac{2d}{d_{\max}^2-d_F^2}\,\mathrm{d}d.
\label{eq:PS6_trunc}
\end{equation}

\textbf{State $S_7$} corresponds to correctly classification of FF UT with conservative rate:
\[
S_7 = \{\, d\ge d_F,~ \hat{d}\ge d_F,~ \hat{d}\geq d\,\}.
\]
Hence,
\begin{align}
\Pr(S_7|d)
= &= \Pr(\hat d \ge d_F \mid d,\,\hat d\ge0) \\
&= \frac{\Pr(\hat d \ge d_F \mid d)}{\Pr(\hat d\ge0\mid d)}
= \frac{Q\!\big(\tfrac{d_F-d}{\sigma_d}\big)}{\Phi\!\big(\tfrac{d}{\sigma_d}\big)}. \nonumber
\label{eq:S7_cond_trunc},
\end{align}
and
\begin{equation}
P_{S_7}
=
\frac{1}{4} \int_{d_F}^{d_{\max}}
\frac{Q\!\big(\tfrac{d_F-d}{\sigma_d}\big)}
     {\Phi\!\big(\tfrac{d}{\sigma_d}\big)}
\;\frac{2d}{d_{\max}^2-d_{F}^2}\,\mathrm{d}d.
\label{eq:S7_uncond_trunc_total}
\end{equation}

\textbf{State $S_8$} corresponds to correctly classification of FF UT but in outage due to rate overestimation:
\[
S_8 = \{\, d\ge d_F,~ \hat{d}\ge d_F,~ \hat{d}<d\,\}.
\]
Hence,
\begin{equation}
\Pr(S_8|d)
= \frac{\Pr(d_F\leq \hat d < d \mid d)}{\Pr(\hat d\ge0\mid d)}
=\frac{\tfrac{1}{2}-\Phi\!\big(\tfrac{d_F-d}{\sigma_d}\big)}%
       {\Phi\!\big(\tfrac{d}{\sigma_d}\big)}
\end{equation}
and
\begin{equation}
P_{S_8}
  = \frac{1}{4} \int_{d_F}^{d_{\max}}
  \frac{\tfrac{1}{2}-\Phi\!\big(\tfrac{d_F-d}{\sigma_d}\big)}%
       {\Phi\!\big(\tfrac{d}{\sigma_d}\big)}
  \;\frac{2d}{d_{\max}^2-d_F^2}\,\mathrm{d}d.
\end{equation}
\color{black}
Finally, all probabilities are normalized so that 
$\sum_{i=1}^{8} P_{S_i}=1$,
and the $8\times8$ transition matrix is
\[
\mathbf{P} =
\begin{bmatrix}
P_{\mathrm{S1}} & P_{\mathrm{S2}} & \cdots & P_{\mathrm{S8}}\\
\vdots & \vdots & \ddots & \vdots\\
P_{\mathrm{S1}} & P_{\mathrm{S2}} & \cdots & P_{\mathrm{S8}}
\end{bmatrix}.
\]

\subsection{Computing MGFs}

In \textbf{State~1} ($H_0|H_0$), the link truly operates in the NF region ($d<d_F$) and is also estimated as NF ($\hat{d}<d_F$). The instantaneous service rate is determined by the continuous-aperture Shannon capacity (Eq.~\ref{eq:near_capacity}), evaluated at the estimated distance $\hat{d}$. 
State~1 therefore implies $d \le \hat{d} < d_F$.

The MGF of the service process conditional on State~1 is
\begin{equation}
M_{R|S1}(\theta)\mid d
= \frac{1}{\sqrt{2\pi}\sigma_d}
  \int_{d}^{d_F}
  \exp\!\big[-\theta R(\hat{d})\big]\,
  e^{-\frac{(\hat{d}-d)^2}{2\sigma_d^2}}
  d\hat{d}.
\label{eq:MGF_S1_exact}
\end{equation}

In general, the MGFs for all states can be expressed as
\begin{equation}
\begin{split}
M_{R|S_i}(\theta)\mid d
= & \mathbb{E}\!\left[e^{-\theta R(\hat{d})}\mid S_i\right]\\
= & \frac{1}{\sqrt{2\pi}\sigma_d}
  \int_{\mathcal{D}_i}
  e^{-\theta R(\hat{d})}\, 
  e^{-\frac{(\hat{d}-d)^2}{2\sigma_d^2}}\,d\hat{d},
  \end{split}
\end{equation}
where $\mathcal{D}_i = [d, d_F]$ for $i=1,\ldots,4$ and $\mathcal{D}_i = [d, d_{\max}]$ for $i=5,\ldots,8$ denote the corresponding decision regions.  

It is important to note that, for certain states (specifically \textbf{States~2, 4, 5, 6, and~8}), the estimated rate exceeds the Shannon capacity, leading to zero effective service rate. Consequently, the MGFs for these states equal~1, i.e., $M_{R|S_i}(\theta)=1$ for $i\in\{2, 4, 6, 8\}$.  
Moreover, States~1 and~3 share the same MGF expression, as do States~5 and~7.

The unconditional MGF of the service process for \textbf{States~1 and ~3 } is given by
\begin{equation}
\begin{split}
M_{R|S_{1,3}}(\theta)
= &
  \frac{\sqrt{2}}{\pi\sigma_d(d_F^2 - d_{\min}^2)} \\
  & \int_{d_{\min}}^{d_F}
  \int_{d}^{d_F}
  d\, e^{-\theta R(\hat{d})}\, 
  e^{-\frac{(\hat{d}-d)^2}{2\sigma_d^2}}
  d\hat{d}\,dd.
  \end{split}
\label{eq:MGF_uncond_S13}
\end{equation}

Similarly, the unconditional MGF for \textbf{State~7} is
\begin{equation}
\begin{split}
M_{R|S_{7}}(\theta)
= & \frac{1}{\sqrt{2\pi}\sigma_d}
  \frac{2}{d_{\max}^2 - d_F^2} .\\
  & \int_{d_F}^{d_{\max}}
  \int_{d}^{d_{\max}}
  d\, e^{-\theta R(\hat{d})}\, 
  e^{-\frac{(\hat{d}-d)^2}{2\sigma_d^2}}
  d\hat{d}\,dd.
  \end{split}
\label{eq:MGF_uncond_S57}
\end{equation}

% Finally, the MGF for \textbf{State~4} can be expressed as
% \begin{equation}
% M_{R|S_4}(\theta)
% = \frac{1}{\sqrt{2\pi}\sigma_d}
%   \frac{2}{d_F^2 - d_{\min}^2}
%   \int_{d_{\min}}^{d_F}
%   \int_{0}^{d}
%   d\, e^{-\theta C_f(\hat{d})}\, 
%   e^{-\frac{(\hat{d}-d)^2}{2\sigma_d^2}}
%   d\hat{d}\,dd.
% \label{eq:MGF_uncond_S4}
% \end{equation}

Once all MGFs are obtained, the diagonal MGF matrix $\mathbf{\Theta}(\theta)$ can be constructed as
\begin{equation}
\mathbf{\Theta}(\theta)
= \mathrm{diag}\!\big(M_{R|S_1}(\theta),\,M_{R|S_2}(\theta),\,\ldots,\,M_{R|S_8}(\theta)\big).
\end{equation}

\subsection{Final Expression of EC}

At a given time, the system operates in one of the eight possible states, forming a Markov chain process. Accordingly, the effective capacity (EC) defined in Eq.~\ref{eq:ECStandard} can be alternatively expressed as \cite{Chang:TNCS:2012}
\[
\text{EC} = -\frac{1}{\theta} \log_e \, \text{sp}(\mathbf{P}\mathbf{\Theta}(\theta)),
\]
where $\text{sp}(\cdot)$ denotes the spectral radius of a matrix, representing the largest absolute value among its eigenvalues. This formulation characterizes the maximum constant arrival rate that the system can support under a given delay exponent $\theta$. 

By considering the transition probabilities and the corresponding MGFs of all states, the closed-form tractable expression for EC is given as
\[
\text{EC} = -\frac{1}{\theta} \ln{\left(\sum_{i=1}^{8} P_{\mathrm{S_i}}\, M_{R|S_i}(\theta)\right)}.
\]

\section{Simulations Results}
We now present the numerical results to investigate the impact of different system parameters on the performance of EC. Unless otherwise stated, simulations are performed using a carrier frequency of  $f = 28~\text{GHz}$ (corresponding to a wavelength $\lambda = 10.7~\text{mm}$). The BS and UT employ continuous-aperture ULAs 
of lengths $L_t = 100\lambda$ and $L_r = 25\lambda$, respectively, 
yielding a Fraunhofer distance as $d_F = 2L_tL_r/\lambda$. 
The UT distance is uniformly distributed in $[0,500]~\text{m}$, 
while the distance estimation error variance is varied as 
$\sigma_d \in [0,~20]~\text{m}$. The Tx power is set to $P = 1~\text{W}$.

\begin{figure}[htb!]
    \centering
    \includegraphics[width=0.9\linewidth]{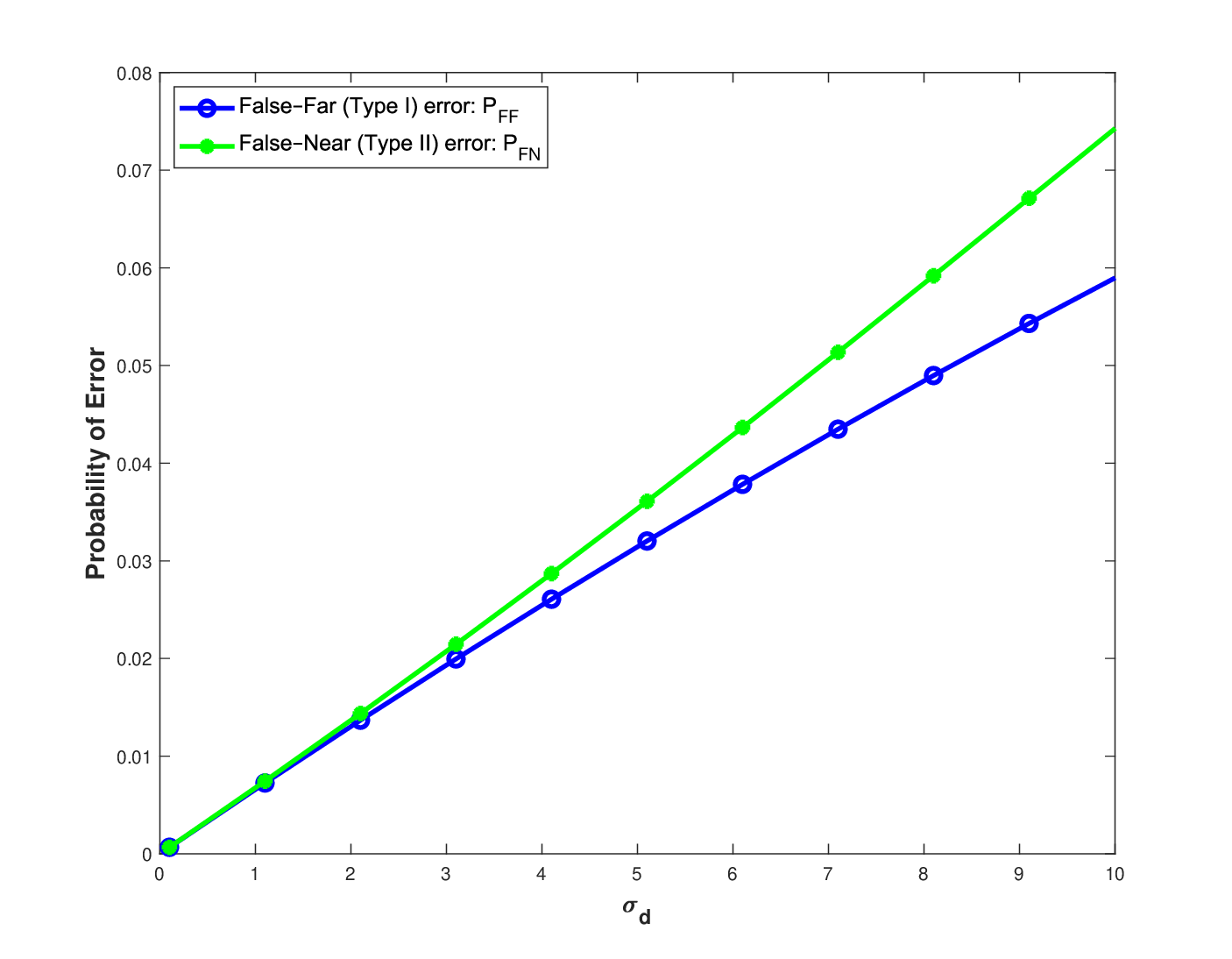}
    \caption{Probability of error vs. estimation uncertainty $\sigma_d$ }
    \label{fig:Prob_FA}
\end{figure}  
As shown in Fig.~\ref{fig:Prob_FA}, both false-far ($P_{\mathrm{FF}}$) and false-near ($P_{\mathrm{FN}}$) probabilities increase monotonically with the estimation uncertainty $\sigma_d$. Larger estimation uncertainty leads to higher misclassification of the propagation regime. The NF regime shows slightly lower error rates due to stronger geometric coupling and higher signal strength, while FF users are more susceptible to incorrect classification. This highlights the crucial role of accurate distance estimation for near–far field boundary decisions in large-aperture 6G systems.
 
\begin{figure}[htb!]
    \centering
    \includegraphics[width=0.9\linewidth]{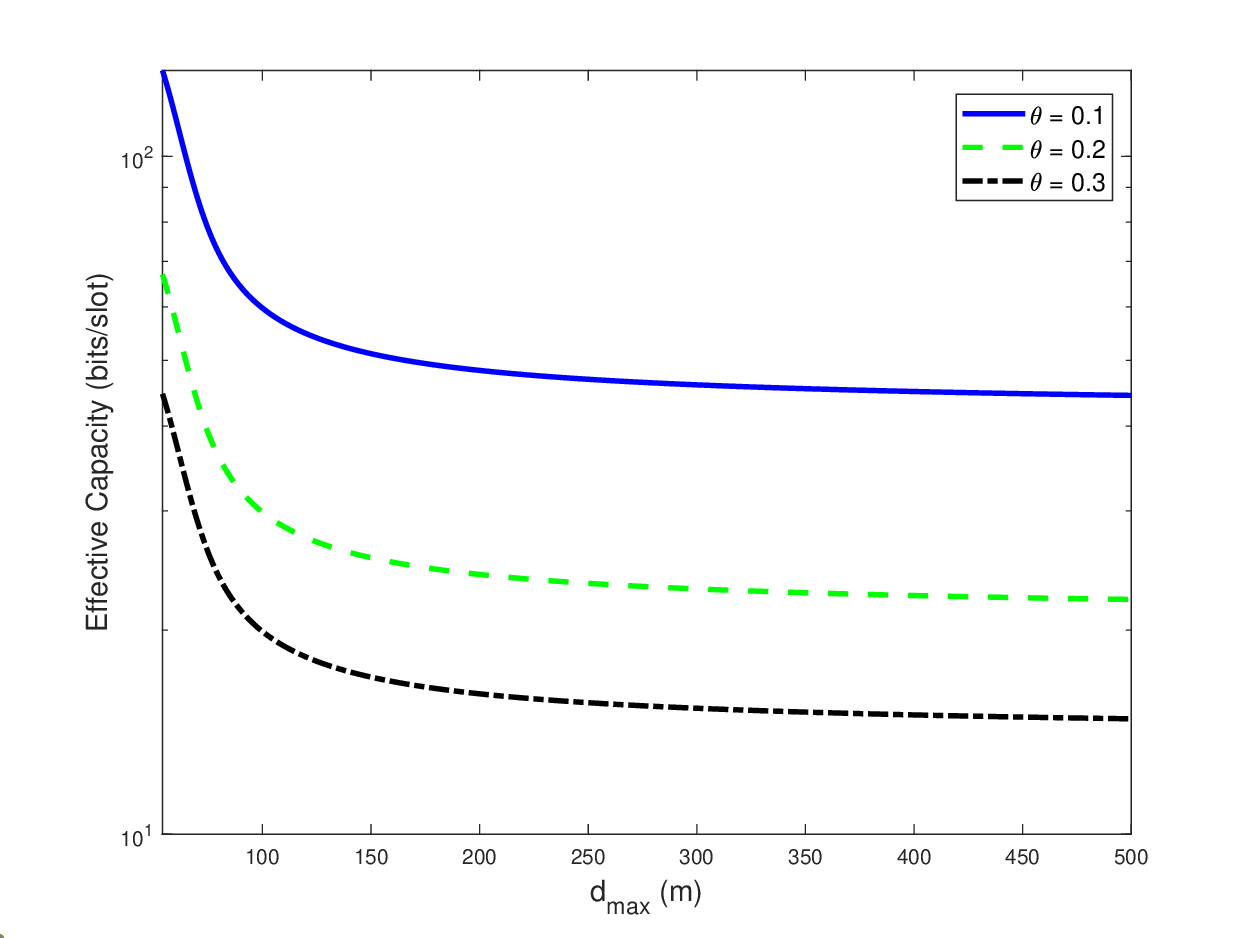}
    \caption{EC vs. boundary of FF ($d_{\max}$) from BS  }
    \label{fig:d_max_vsEC}
\end{figure}

Fig.~\ref{fig:d_max_vsEC} illustrates that the EC decreases with increase in the FF boundary $d_{\max}$. A larger FF coverage extends the region of low SNR, which lowers the achievable rate under delay constraints. Further, as $d_{\max}$ increases, the likelihood of users operating near the boundary---and hence misclassification---increases, further degrading EC. Thus, stringent delay constraints amplify the impact of rate mismatch and outages.

% As shown in Fig.~\ref{fig:d_min_vsEC}, the EC improves as the minimum distance $d_{\min}$ increases. This behavior occurs because users are on average closer to the base station, experiencing stronger coupling and higher spatial focusing gains. The improvement is nearly linear for small $d_{\min}$ and tends to saturate as $d_{\min}$ increases further. This confirms that dense NF operation enhances delay-limited throughput performance.
% \begin{figure}
%     \centering
%     \includegraphics[width=0.9\linewidth]{figures/d_1_VsEC_Truncated.eps}
%     \caption{EC vs. Near-Field boundary.}
%     \label{fig:d_min_vsEC}
% \end{figure}

\begin{figure}
    \centering
    \includegraphics[width=0.9\linewidth]{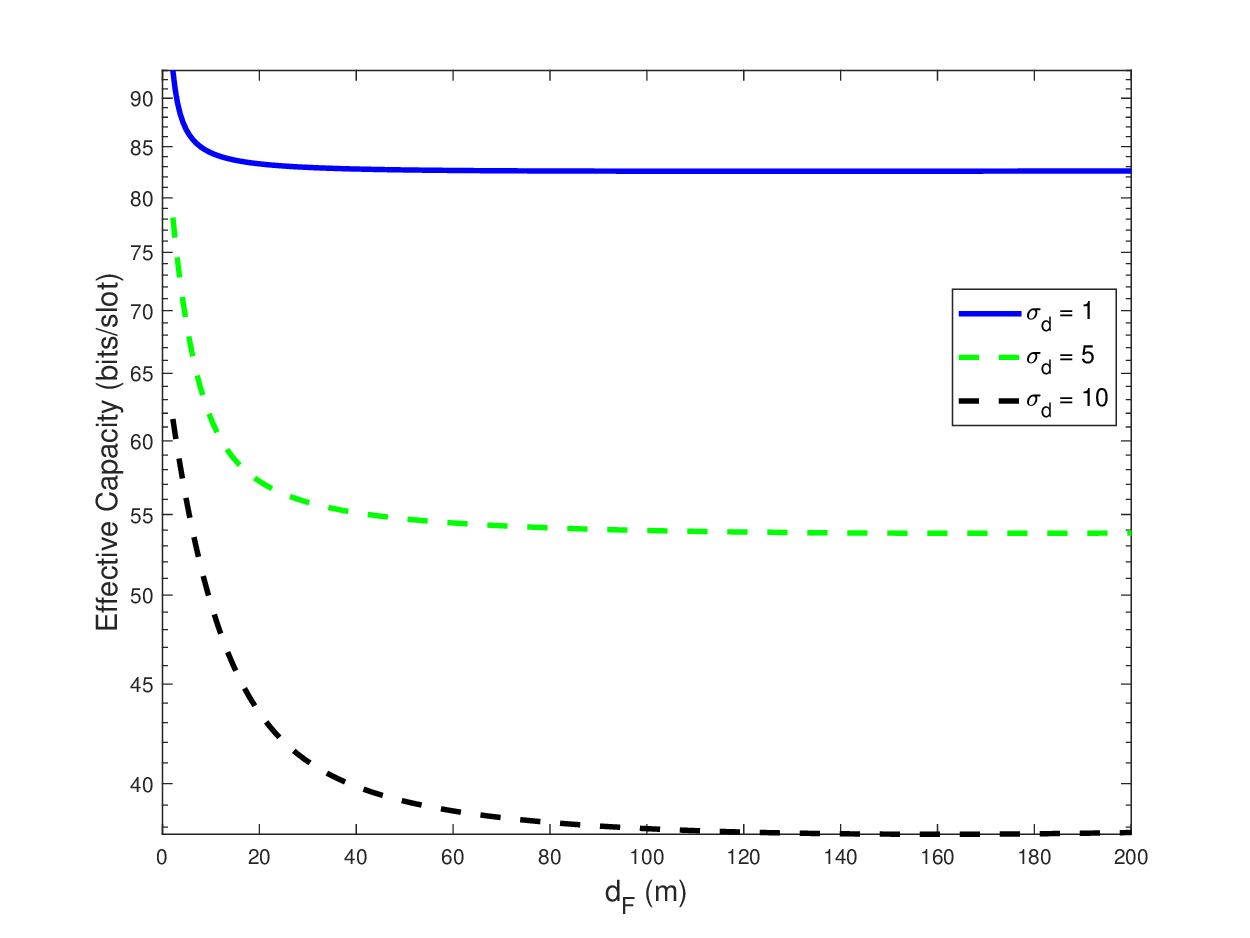}
    \caption{EC vs. boundary of NF ($d_F$) from BS}
    \label{fig:d_F_vsEC}
\end{figure}

Fig.~\ref{fig:d_F_vsEC} shows EC for different
values of effective degree of freedom (EDOF) based field
boundary ($d_F$). EC decreases monotonically with increasing  \( d_F \) for all values of the estimation uncertainty \( \sigma_d \). This behavior occurs because a larger \( d_F \) expands the NF region, forcing the scheduler to use the NF capacity model even when the user is farther away, thereby reducing the achievable rate. Additionally, as \( d_F \) increases, the average channel gain weakens due to greater propagation loss and reduced array gain efficiency. The degradation is more pronounced for higher estimation uncertainty (\( \sigma_d = 10 \)), where regime misclassification becomes dominant. In contrast, for smaller \( \sigma_d \), the EC remains higher and relatively stable, reflecting improved robustness to boundary estimation errors.

\begin{figure}
    \centering
    \includegraphics[width=0.9\linewidth]{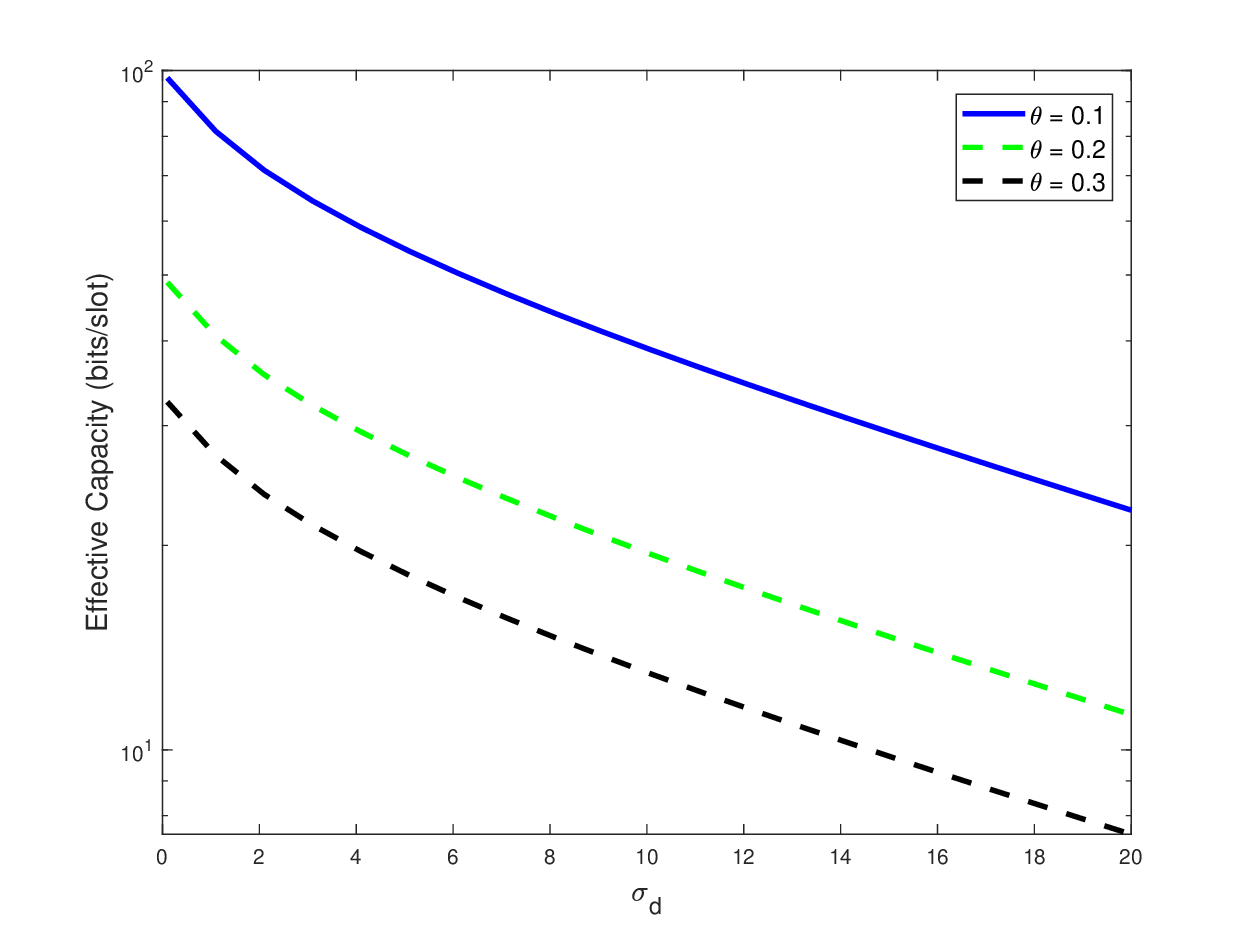}
    \caption{EC vs. distance estimation uncertainty ($\sigma_d$)}
    \label{fig:NoisevsEC}
\end{figure}

Fig.~\ref{fig:NoisevsEC} shows that EC degrades significantly with increasing $\sigma_d$. A larger estimation error leads to more frequent transmission rate mismatches, causing higher outage probability and degraded QoS performance. The degradation becomes sharper for higher $\theta$, as stricter delay requirements limit the tolerance to channel uncertainty. This emphasizes the importance of precise range estimation (e.g., ToA-based) to maintain high reliability and throughput under delay constraints.
  
\section{Conclusion}

This study established the EC analytical framework to characterize the delay-sensitive performance of emerging URLLC 6G networks that will utilize both near-field and far-field communication regimes. By jointly modeling these distinct propagation regions and incorporating distance estimation uncertainty, we derived a tractable closed-form expression for the EC, allowing for precise quantification of the network's QoS guarantees. Our numerical evaluation provided crucial insights into how system parameters, specifically distance estimation variance, the QoS exponent, and the Fraunhofer distance, fundamentally impact the achievable EC.

%\section*{Acknowledgment}

\appendix
\subsection*{ToA-based Distance Estimation and CRLB Analysis}

In a two-way ranging configuration, the BS transmits a known pilot
signal $x(t)$, which is reflected or re-transmitted by the UT and received back at the BS.
The received echo signal at the BS is modeled as
\begin{equation}
    y_{\text{BS}}(t) = \beta\, x(t - \tau_{\text{rt}}) + n_{\text{BS}}(t),
    \label{eq:rx_echo}
\end{equation}
where $\beta$ denotes the complex round-trip channel coefficient and
$\tau_{\text{rt}} = \frac{2d}{c}$ represents the round-trip propagation delay,
with $d$ the true distance and $c$ the propagation speed.
The BS estimates the ToA by correlating the Tx and Rx signals:
\begin{equation}
    r_{xy}(\tau') = \int y_{\text{BS}}(t)\, x^*(t - \tau')\, dt,
\end{equation}
and the delay estimate is obtained as
\begin{equation}
    \hat{\tau}_{\text{rt}} = \arg\max_{\tau'} |r_{xy}(\tau')|.
    \label{eq:toa_est}
\end{equation}
The estimated distance is given as follows
\begin{equation}
    \hat{d} = \frac{c\,\hat{\tau}_{\text{rt}}}{2}.
    \label{eq:range_est}
\end{equation}

\subsubsection*{Fisher Information and CRLB for ToA}
Let $\tau$ denote the unknown ToA parameter to be estimated from~\eqref{eq:rx_echo}.
Assuming $x(t)$ is deterministic, known, and of unit energy
($\int |x(t)|^2 dt = 1$), and that the noise $n_{\text{BS}}(t)$ is complex AWGN with variance $\sigma_n^2$,
the Fisher information for $\tau$ is given by~\cite{fundamentals}
\begin{equation}
    \mathcal{I}(\tau)
    = \frac{2|\beta|^2}{\sigma_n^2}
      \int_{-\infty}^{\infty}
      \left|
      \frac{\partial x(t-\tau)}{\partial \tau}
      \right|^2 dt.
\end{equation}
Since $    \frac{\partial x(t-\tau)}{\partial \tau}
    = -\dot{x}(t-\tau)$,
where $\dot{x}(t-\tau) = \tfrac{d}{dt}x(t-\tau)$ denotes the time derivative of the transmitted waveform evaluated at time $t-\tau$, the Fisher information becomes
\begin{equation}
    \mathcal{I}(\tau)
    = \frac{2|\beta|^2}{\sigma_n^2}
      \int_{-\infty}^{\infty} |\dot{x}(t-\tau)|^2 dt.
\end{equation}
Using Parseval’s theorem,
$\int |\dot{x}(t)|^2 dt = (2\pi)^2 \int f^2 |X(f)|^2 df$, 
the Cramér–Rao lower bound (CRLB) on the ToA estimate is
\begin{equation}
    \mathrm{Var}(\hat{\tau}) \ge
    \frac{1}{8\pi^2\,\gamma\,\beta_2},
    \label{eq:crlb_tau}
\end{equation}
where $
    \beta_2 = \int f^2 |X(f)|^2 df$,
$    \gamma = \frac{|\beta|^2}{\sigma_n^2}$
By propagation through~\eqref{eq:range_est}, the CRLB for the distance estimate becomes
\begin{equation}
    \mathrm{Var}(\hat{d})
    = \left(\frac{c}{2}\right)^2 \times \mathrm{Var}(\hat{\tau})
    \ge
    \frac{c^2}{32\pi^2\,\gamma\,\beta_2}.
    \label{eq:crlb_range}
\end{equation}

\subsubsection*{Extension to Continuous Transmit and Receive Apertures}

When both the base station and the UT terminal employ continuous apertures
of lengths $L_t$ and $L_r$, respectively, the received echo field at the BS
can be modeled as
\[
y(x_r,t) =
\int_{-L_t/2}^{L_t/2}
\beta(x_t,x_r)\,
x(t - \tau(x_t,x_r))\,dx_t + n(x_r,t),
\]
where $\tau(x_t,x_r) = \frac{2}{c}\sqrt{d^2 + (x_t - x_r)^2}$ denotes the
two-way delay between aperture points $x_t$ and $x_r$.
The corresponding Fisher information for the ToA is
\[
\mathcal{I}^{\text{CA}}(\tau)
= \frac{2|\beta|^2}{\sigma_n^2}
\int_{-L_t/2}^{L_t/2}
\int_{-L_r/2}^{L_r/2}
\int_{-\infty}^{\infty}
\left|
\frac{\partial x(t-\tau(x_t,x_r))}{\partial \tau}
\right|^2
dt\,dx_t\,dx_r.
\]
Under the narrowband assumption, this simplifies to
$
\mathcal{I}^{\text{CA}}(\tau)
\approx
\frac{2|\beta|^2 L_t L_r}{\sigma_n^2}
\int_{-\infty}^{\infty} |\dot{x}(t)|^2 dt,
$
yielding the CRLB
\[
\mathrm{Var}^{\text{CA}}(\hat{\tau})
\ge
\frac{1}{
8\pi^2\,L_t L_r\,\gamma,\beta_2},
\qquad
\sigma_d^2
\ge
\frac{c^2}{
32\pi^2\,L_t L_r\,\gamma,\beta_2}.
\]

\footnotesize{
\bibliographystyle{IEEEtran}
\bibliography{refs}
}

\end{document}